\documentclass[aps,twocolumn,floatfix,amsmath,longbibliography,amsfonts]{revtex4-1}

\usepackage{graphicx}
\usepackage{braket}
\usepackage{bm}
\usepackage{xcolor}
\usepackage{upgreek}
\usepackage{tabularx}
\usepackage{tikz}
\usepackage{todonotes}
\usepackage[normalem]{ulem}
\usepackage[utf8]{inputenc}

\usepackage{hyperref}

\begin{document}
\title{Exchange interaction in the yellow exciton series of cuprous oxide}
\author{Patric Rommel}
\email[Email: ]{patric.rommel@itp1.uni-stuttgart.de}
\author{J\"org Main}
\affiliation{Institut f\"ur Theoretische Physik 1, Universit\"at
  Stuttgart, 70550 Stuttgart, Germany}
\author{Andreas Farenbruch$^1$}
\author{Dmitri R. Yakovlev$^{1,2}$}
\author{Manfred Bayer$^{1,2}$}
\affiliation{$^1$Experimentelle Physik 2, Technische Universität Dortmund, 44221 Dortmund, Germany\\ $^2$Ioffe Institute, Russian Academy of Sciences, 194021 St.\ Petersburg, Russia}

\date{\today}

\begin{abstract}
We experimentally and numerically investigate the exchange interaction
of the yellow excitons in cuprous oxide.
By varying the material parameters in the numerical calculations, we
can interpret experimental findings and understand their origin in the
complex band structure and central-cell corrections.
In particular, we experimentally observe the reversal of the ortho- and paraexciton
for the $2S$ yellow exciton, and explain this phenomenon by an avoided
crossing with the green $1S$ orthoexciton in a detailed numerical analysis.
Furthermore, we discuss the exchange splitting as a function of the
principal quantum number $n$ and its deviation from the
$n^{-3}$ behavior expected from a hydrogenlike model.
We also explain why the observed exchange splitting of the green $1S$
exciton is more than twice the splitting of the yellow $1S$ state.
\end{abstract}

\maketitle

\section{Introduction}
\label{sec:introduction}
The yellow exciton series in cuprous oxide has been shown to closely
match a hydrogenlike system in many respects~\cite{GiantRydbergExcitons}.
Still, there are a number of characteristic effects of the complex
band structure.
For example, a fine structure splitting between $P$ and $F$ states can be
observed~\cite{AszmannGUE2016}.
In the case of small radii, additional central-cell corrections to the
valence band Hamiltonian
have to be added to achieve a satisfactory description~\cite{frankevenexcitonseries}.
Due to the cubic symmetry of the crystal, the angular momentum is not
a good quantum number anymore, and thus the $S$ states with small
extension are also coupled to other angular momenta, such as the $D$
states.
This is especially important when considering the green $1S$ state,
which lies in between the yellow spectrum.
Because of this coupling, the central-cell corrections
also affect the energetically higher lying states of the
yellow series.

The exchange interaction, which is part of the central-cell corrections, causes a characteristic splitting between states depending on the
relative alignment of the electron and hole spins, i.e., between the spin-singlet and spin-triplet states.
The spin-triplet dark exciton states have been proposed for use in
quantum computational applications and for the possible realization of a Bose-Einstein condensate~\cite{Poem2010,Snoke_2014,Beian_2017}.
The dark paraexciton series in cuprous oxide is not accessible by electric
dipole and quadrupole absorption experiments, because the paraexcitons
have no spin-singlet component and are therefore spin-flip forbidden
to all orders in electric transitions.
This selection rule can be circumvented using stress~\cite{Kuwabara1977} or by application of an external
magnetic field which leads to a mixing between spin-singlet and
triplet states~\cite{Brandt2007,Mysyrowicz1983}. We use magnetic-field-induced Second Harmonic Generation (SHG)
spectroscopy for the measurements of the paraexcitons, extrapolating their magnetic-field dependent energies to
zero field strength.

A naive treatment of the exchange interaction leads to the expectation
that the orthoexciton is shifted to higher energies than the paraexciton, a result in line with Hund's rule.
Numerical calculations have shown that this expectation is
contradicted in the case of the yellow $2S$ exciton
state in cuprous oxide~\cite{frankevenexcitonseries}.
This has now been confirmed in experiments by
Farenbruch \emph{et al.}~\cite{Farenbruch2020Dark}.
They explain this by appealing to the influence of the
green $1S$ state.
In this paper, we discuss and confirm this explanation in greater detail, using the possibility of changing the material parameters
in the numerical simulations to study the spectrum in experimentally inaccessible ways.
{Going beyond the discussion in Ref.~\cite{Farenbruch2020Dark}, we present the precise mechanism responsible
for the reversed energies of the $2S$ ortho- and paraexcitons.}
We then study the exchange splitting as a function of the principal quantum number $n$. Based on a hydrogenlike calculation, the 
splitting is expected to decrease with $n^{-3}$. For the yellow excitons in cuprous oxide, there are deviations from this.
We numerically investigate the origin of these deviations. We also explain why the exchange splitting of the green $1S$
state is more than twice the exchange splitting of the yellow $1S$ state.
{To the best of our knowledge, this represents the first detailed discussion of these features of the
dark exciton series in cuprous oxide.}

The paper is organized as follows: We first introduce the Hamiltonian
and focus on the central-cell corrections in
Sec.~\ref{subsec:Hamiltonian}.
We briefly explain our numerical methods in Sec.~\ref{subsec:Numerical}
and our experimental methods in Sec.~\ref{subsec:experiment}.
In Sec.~\ref{sec:ResultsDiscussion}, we discuss the reversal between
the $2S$ ortho- and paraexcitons (Sec.~\ref{subsec:2S}), the exchange
splitting as a function of the principal quantum number
(Sec.~\ref{subsec:PrincipalQuantumNumber}) and the splitting of the
yellow $1S$ state versus the green $1S$ state (Sec.~\ref{subsec:YellowVsGreen}).
We finally conclude in Sec.~\ref{sec:Conclusion}.

\section{Methods and Materials}
\label{sec:MethodsMaterials}
In this section, we introduce the theoretical background for the
description of the yellow exciton series and our method of numerical
diagonalization.
Furthermore, we explain our experimental methods.

\subsection{Hamiltonian including central-cell corrections}
\label{subsec:Hamiltonian}

{Excitons are described as hydrogenlike excitations of the crystal,
where an electron is lifted from one of the valence bands to the
conduction band, leaving behind a hole.
The complex valence band structure can be described via a quasispin
$\boldsymbol{I}$ in addition to the hole spin
$\boldsymbol{S}_\mathrm{h}$.
This description introduces additional degrees of freedom compared to
the hydrogenlike model.
For a detailed analysis, especially of the ortho and para 2S states,
central-cell corrections are required.
Details of derivations are already presented in the literature
\cite{ImpactValence,frankevenexcitonseries,knox1963excitons,Uihlein1981,Froehlich1979,KavoulakisBohrRadius}.
For the convenience of the reader in this paper we briefly
recapitulate the basic equations.}

{The yellow and green exciton series in cuprous oxide (Cu$_2$O) belong
to the $\Gamma_7^+$ and $\Gamma_8^+$ valence band, respectively.
They can be described}
using the Hamiltonian~\cite{ImpactValence}
\begin{equation}
  H = E_{\rm g}+H_{\rm e}(\boldsymbol{p}_{\rm e})+H_{\rm h}(\boldsymbol{p}_{\rm h})
  +V(\boldsymbol{r}_{\rm e}-\boldsymbol{r}_{\rm h}) + V_{\mathrm{CCC}}^{\mathrm{H}}\!\left(\boldsymbol{r}\right) \, .
\label{eq:hamiltonian}
\end{equation}
Here $E_{\rm g}$ is the band gap between the uppermost $\Gamma_7^+$ valence band and the lowermost
$\Gamma_6^+$ conduction band. The kinetic energies of electron and hole are given by
\begin{align}
  H_{\rm e}(\boldsymbol{p}_{\rm e}) &= \frac{\boldsymbol{p}_{\rm e}^2}{2m_{\rm e}} \, , \\
  H_{\rm h}(\boldsymbol{p}_{\rm h}) &= H_{\rm SO}+\frac{1}{2\hbar^2m_0}
  \{\hbar^2(\gamma_1+4\gamma_2)\boldsymbol{p}^2_{\rm h}\nonumber\\
&+2(\eta_1+2\eta_2)\boldsymbol{p}^2_{\rm h}(\boldsymbol{I}\cdot\boldsymbol{S}_{\rm h})\nonumber\phantom{\frac{1}{2}}\\
&-6\gamma_2(p^2_{\rm h1}\boldsymbol{I}^2_1+{\rm c.p.})
-12\eta_2(p^2_{\rm h1}\boldsymbol{I}_1\boldsymbol{S}_{\rm h1}+{\rm c.p.})\nonumber\phantom{\frac{1}{2}}\\
&-12\gamma_3(\{p_{\rm h1},p_{\rm h2}\}\{\boldsymbol{I}_1,\boldsymbol{I}_2\}+{\rm c.p.})\nonumber\phantom{\frac{1}{2}}\\
\phantom{\frac{1}{2}} &-12\eta_3(\{p_{\rm h1},p_{\rm h2}\}(\boldsymbol{I}_1\boldsymbol{S}_{\rm h2}
         +\boldsymbol{I}_2\boldsymbol{S}_{\rm h1})+{\rm c.p.})\} \, .
         \label{eq:ElectronHoleKinetic}
\end{align}
{We use} the electron mass in the crystal $m_\mathrm{e}$ and in vacuum $m_0$,
 the Luttinger parameters $\gamma_i$, $\eta_i$,
{the spin $\boldsymbol{S}_{\rm h}$ and quasispin $\boldsymbol{I}_i$ of the hole,}
the momenta
$\boldsymbol{p}_\mathrm{e}$ and $\boldsymbol{p}_\mathrm{h}$ of the
electron and hole, respectively. {The indices $i = 1,2,3$ for the momenta, positions, quasispin and hole
spin denote the Cartesian $x$, $y$, and $z$-components, ``c.p.'' denotes cyclic permutation.}
{The spin-orbit coupling term reads}
\begin{equation}
  H_{\rm SO}=\frac{2}{3}\Delta
  \left(1+\frac{1}{\hbar^2}\boldsymbol{I}\cdot\boldsymbol{S}_{\rm h}\right)\,.
  \label{eq:SOcoupling}
\end{equation}
Electron and hole interact via the screened Coulomb potential
\begin{equation}
  V(\boldsymbol{r}_{\rm e}-\boldsymbol{r}_{\rm h}) =
  -\frac{e^2}{4\pi\varepsilon_0\varepsilon|\boldsymbol{r}_{\rm e}-\boldsymbol{r}_{\rm h}|}\,,
\label{eq:Coulomb_pot}
\end{equation}
with the dielectric constant $\varepsilon = \varepsilon_{\mathrm{s}1}$ and the positions of the electron $\boldsymbol{r}_{\rm e}$ and hole $\boldsymbol{r}_{\rm h}$.
We express the system in relative and center-of-mass coordinates~\cite{Schmelcher1992},
\begin{eqnarray}
  \boldsymbol{r}&= \boldsymbol{r}_{\rm e}-\boldsymbol{r}_{\rm h}\, ,\quad
  \boldsymbol{R}=\frac{m_{\rm h}\boldsymbol{r}_{\rm h}+m_{\rm e}\boldsymbol{r}_{\rm e}}{m_{\rm h}+m_{\rm e}}\, ,\nonumber\\
  \boldsymbol{P}&= \boldsymbol{p}_{\rm e}+\boldsymbol{p}_{\rm h}\, ,\quad
  \boldsymbol{p}=\frac{m_{\rm h}\boldsymbol{p}_{\rm e}-m_{\rm e}\boldsymbol{p}_{\rm h}}{m_{\rm h}+m_{\rm e}} \, ,
\end{eqnarray}
with vanishing center-of-mass momentum $\boldsymbol{P} = 0$.

For small separations, the electron-hole pair probes features of the
crystal structure not captured by the valence band terms given 
in Eq.~\eqref{eq:ElectronHoleKinetic}.
{Furthermore, the dielectric constant $\varepsilon$ in Eq.~\eqref{eq:Coulomb_pot}
is no longer constant.}
The corresponding corrections to the Hamiltonian are the central-cell
corrections~\cite{knox1963excitons,Uihlein1981,Froehlich1979,KavoulakisBohrRadius}, {and they primarily concern
the states with principal quantum number $n \leq 2$.}
As derived in Ref.~\cite{frankevenexcitonseries}, the central-cell corrections in cuprous oxide are given by
\begin{equation}
V_{\mathrm{CCC}}^{\mathrm{H}}\!\left(\boldsymbol{r}\right) =  V^{\mathrm{H}} + H_\mathrm{exch} + V_d\,,
\label{eq:CCCHaken}
\end{equation}
where
\begin{align}
V^{\mathrm{H}} = -\frac{e^{2}}{4\pi\varepsilon_{0}r}
\Bigg[\frac{1}{2\varepsilon_{1}^{*}}\left(e^{-r/\rho_{\mathrm{h}1}}+e^{-r/\rho_{\mathrm{e}1}}\right)\nonumber\\
+\frac{1}{2\varepsilon_{2}^{*}}\left(e^{-r/\rho_{\mathrm{h}2}}+e^{-r/\rho_{\mathrm{e}2}}\right)\Bigg]\,,
\label{eq:Haken}
\end{align}
is the Haken potential describing the modification of the dielectric constant
{for electron-hole separations on the order of the polaron radii,}
with further corrections modeled by a contact interaction
\begin{equation}
V_d = -V_0 V_{\mathrm{uc}}\delta\!\left(\boldsymbol{r}\right)\,,
\label{eq:Correction}
\end{equation}
with the volume of a lattice unit cell $V_\mathrm{uc} = a^3$ and the lattice constant $a$.
We use
\begin{equation}
\frac{1}{\varepsilon_i^\ast} = \frac{1}{\varepsilon_{\mathrm{b}i}} - \frac{1}{\varepsilon_{\mathrm{s}i}}\,,
\end{equation}
and the polaron radii
\begin{equation}
\rho_{\mathrm{e}i}=\sqrt{\frac{\hbar}{2m_{\mathrm{e}}\omega_{\mathrm{LO}i}}},\qquad\rho_{\mathrm{h}i}=\sqrt{\frac{\hbar\gamma_1}{2m_0\omega_{\mathrm{LO}i}}},
\end{equation}
with the energies $\hbar\omega_{\mathrm{LO}i}$ of the longitudinal $\Gamma_4^-$ phonons. The relevant phonon branches are marked with $i = 1,2$.
{The exchange interaction is given in
  Ref.~\cite{frankevenexcitonseries} and reads}
\begin{align}
H_\mathrm{exch} &= J_{0}\left(\frac{1}{4}
-\frac{1}{\hbar^{2}}\boldsymbol{S}_{\mathrm{e}}\cdot\boldsymbol{S}_{\mathrm{h}}\right)
 V_{\mathrm{uc}}\delta\!\left(\boldsymbol{r}\right)\nonumber \\
 &= J_{0} \left(1- \frac{1}{2\hbar^{2}}\boldsymbol{S}^2\right) V_{\mathrm{uc}}\delta\!\left(\boldsymbol{r}\right)\,,\label{eq:Exchange}
\end{align}
{where we use the total spin $\boldsymbol{S} = \boldsymbol{S}_{\mathrm{e}} + \boldsymbol{S}_{\mathrm{h}}$ in
the second part of the equation.}
{We want to take a closer look at the exchange interaction
\eqref{eq:Exchange} in the following.\\[1em]}
{We first note, that only $L=0$ states are affected due to the presence of the $\delta$ term.
From the second line in Eq.~\eqref{eq:Exchange} it is clear, that the effect is a lifting of the
states with $S=0$ over the states with $S=1$.}
\begin{figure}
\includegraphics[]{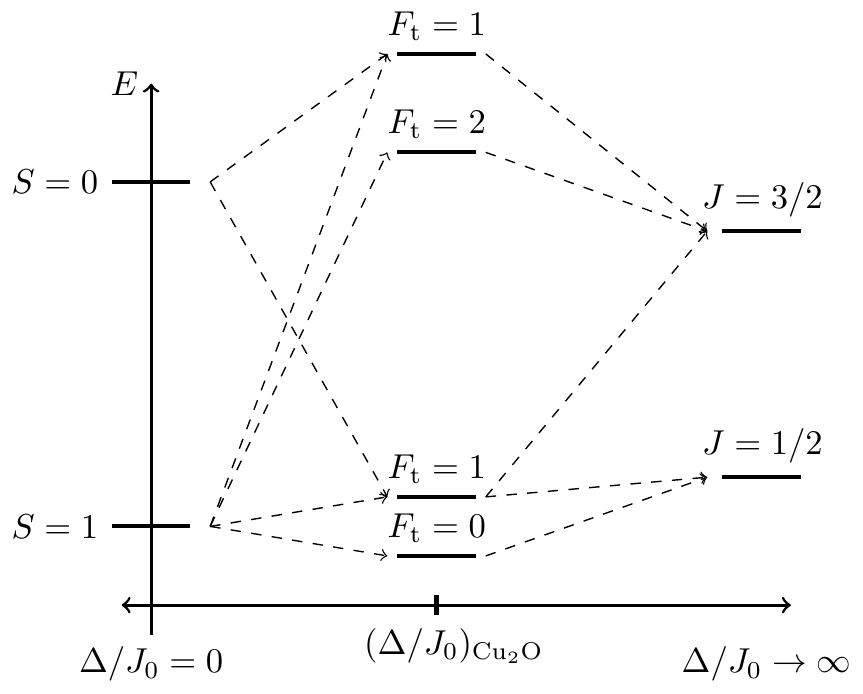}
\caption{Level scheme of $S$ states as a function of the ratio between the exchange interaction strength $J_0$
and the spin-orbit coupling $\Delta$. For $L = 0$, the total angular momentum $F_\mathrm{t}$ can be obtained
by either coupling the quasispin {of the hole} $\boldsymbol{I}$ {with} the total spin
$\boldsymbol{S}=\boldsymbol{S}_\mathrm{e}+\boldsymbol{S}_\mathrm{h}$ or by coupling the
electron spin $\boldsymbol{S}_\mathrm{e}$ {with} the effective hole spin $\boldsymbol{J}=\boldsymbol{I}+\boldsymbol{S}_\mathrm{h}$.
According to Eq.~\eqref{eq:Exchange},
the $S = 0$ singlet states are lifted above the $S = 1$ triplet states for vanishing spin-orbit coupling $\Delta$. For $\Delta/J_0 \rightarrow \infty$,
the splitting between the green $J = 3/2$ and yellow $J=1/2$ states predominates. For general values inbetween, the levels split according to the
total angular momentum $F_\mathrm{t}$. As discussed in Sec.~\ref{subsec:YellowVsGreen} with reference to the matrix element~\eqref{eq:MatrixElementExchange},
for both the green and yellow $S$ states, the $F_\mathrm{t} = 1$ states
are lifted above the $F_\mathrm{t} = 2$ and $F_\mathrm{t} = 0$ states, respectively. 
}
\label{fig:ExchangeVSSO}
\end{figure}
When taking into account the quasispin $\boldsymbol{I}$ in the crystal, $S$ is not a good quantum number anymore,
{and we additionally need to consider the spin-orbit coupling given in Eq.~\eqref{eq:SOcoupling}}.
In Fig.~\ref{fig:ExchangeVSSO} we show the level-scheme of the $S$ states caused by the {competition between the}
exchange interaction and the spin-orbit
coupling as a function of the ratio $\Delta/J_0$.
Note that for the yellow $J=1/2$ states only the threefold degenerate
$F_\mathrm{t} = 1$ orthoexcitons have an $S=0$ component and are therefore dipole allowed.
Here, $\boldsymbol{F}_\mathrm{t} = \boldsymbol{J} + \boldsymbol{S}_\mathrm{e} = \boldsymbol{I} + \boldsymbol{S}$
is the total angular momentum.
While the singlet state $S=0$ is lifted above the triplet state $S=1$,
it is the threefold degenerate $F_\mathrm{t} = 1$ state which is lifted
above the non-degenerate $F_\mathrm{t} = 0$ state, when
only considering the yellow $J=1/2$ states.

We want to perform a quick calculation to understand the behavior of the exchange interaction in a simplified model.
This will allow us to investigate the impact of the correction
terms, i.e., the band structure and central-cell corrections, by
comparing our findings here with the results of the exact numerical
calculation further below.
Evaluating the integral for the matrix elements for the exchange interaction~\eqref{eq:Exchange}
with wavefunctions $\psi_1$ and $\psi_2$
over the $\delta$ term leads to a proportionality
to $\psi_1^\ast(0)\psi_2(0)$. In a hydrogenlike model we choose the $S$ states $\psi_1 = \psi_2 = \psi_{n,L=0,M=0}$. This yields
\begin{align}
&\langle n\,, L=0\,, M=0 |H_\mathrm{exch}| n\,, L=0\,, M=0 \rangle
                \nonumber \\
 &= J_{0}\left(1 -\frac{1}{2\hbar^{2}}\boldsymbol{S}^2\right) V_{\mathrm{uc}} \left(\frac{1}{\pi n a_\mathrm{B}}\right)^3\,, 
 \label{eq:ExchangeFunctionN}
\end{align}
with the Bohr radius $a_\mathrm{B}$.
The relative energetic placement of multiplet states affected by the exchange interaction does not depend on the
principal quantum number $n$ in the hydrogenlike model, but only the strength of the splitting, since $n$ affects only
an overall factor.
As will be shown in Sec.~\ref{subsec:2S}, this does not hold in the case of the yellow excitons in cuprous oxide,
where ortho- and paraexcitons are reversed for
$n=2$. {For the other principal quantum numbers, the order of states is as shown in Fig.~\ref{fig:ExchangeVSSO}.}
Additionally, the splitting decreases with $n^{-3}$ as a function of the principal quantum number in the simplified model.
The situation is more complicated for the yellow exciton series in cuprous oxide, which will be more thoroughly discussed
in Sec.~\ref{subsec:PrincipalQuantumNumber}.

\subsection{Numerical diagonalization}
\label{subsec:Numerical}
To express the Schr\"odinger equation as a generalized eigenvalue problem, we use the complete basis introduced in Ref.~\cite{ImpactValence}.
{In our basis states, the quasispin $\boldsymbol{I}$ and the hole spin $\boldsymbol{S}_\mathrm{h}$ are coupled
to form the effective hole spin $\boldsymbol{J}$.}
Near the $\Gamma$ point, $J$ is an approximate quantum number.
Excitons with $J = 1/2$ and $J = 3/2$ belong to the yellow and green
series, respectively.
We further couple the effective hole spin $\boldsymbol{J}$ and the angular momentum
$\boldsymbol{L}$ to $\boldsymbol{F}$, which is {then} coupled with the electron spin $\boldsymbol{S}_\mathrm{e}$ to the total angular momentum
$\boldsymbol{F}_\mathrm{t}$. The quantization axis is chosen along the [001] direction of the crystal and the corresponding
$z$-component of $\boldsymbol{F}_\mathrm{t}$ is $M_{F_\mathrm{t}}$. For the radial component we use the Coulomb-Sturmian functions~\cite{CoulombSturmForNuclear},
which are rescaled radial hydrogen atom solutions. The total basis states are thus given by
\begin{equation}
 |\Pi \rangle = | N, L, (I, S_\mathrm{h}) J, F, S_\mathrm{e}, F_\mathrm{t}, M_{F_\mathrm{t}} \rangle \,.
 \label{eq:Basis}
\end{equation}
The radial quantum number is defined as $N = n - L - 1$ with the principal quantum number $n$.
The resulting generalized eigenvalue problem is solved using a suitable LAPACK routine~\cite{lapackuserguide3}.
The material parameters used in our calculations are listed in
Table~\ref{tab:MaterialParameters}.
\begin{table}[b]
\renewcommand{\arraystretch}{1.2}
     \centering
     \caption{Material parameters of Cu$_2$O used in the calculations.}
     \begin{tabular}{l|l c}
     \hline
       Energy gap  & $E_{\rm g}=2.17208\,$eV & \cite{GiantRydbergExcitons}\\
       Spin-orbit coupling       & $\Delta=0.131\,$eV &\cite{SchoeneLuttinger}\\
       Effective electron mass  & $m_{\rm e}=0.99m_0$ & \cite{HodbyEffectiveMasses} \\
       Effective hole mass  & $m_{\rm h}=0.58m_0$ & \cite{HodbyEffectiveMasses} \\
       Valence band parameters & $\gamma_1=1.76$&\cite{SchoeneLuttinger}\\
       ~~& $\gamma_2=0.7532$&\cite{SchoeneLuttinger}\\
       ~~& $\gamma_3=-0.3668$&\cite{SchoeneLuttinger}\\
        ~~& $\eta_1=-0.020$&\cite{SchoeneLuttinger}\\
        ~~& $\eta_2=-0.0037$&\cite{SchoeneLuttinger}\\
        ~~& $\eta_3=-0.0337$&\cite{SchoeneLuttinger}\\
       Exchange interaction & $J_0 = 0.792\,\mathrm{eV}$ & \cite{frankevenexcitonseries} \\
       Short distance correction & $V_0 = 0.539\,\mathrm{eV}$ & \cite{frankevenexcitonseries}\\
       Lattice constant & $a=0.42696\,\mathrm{nm}$ & \cite{Swanson1953}\tabularnewline
       Dielectric constants & $\varepsilon_{\mathrm{s}1}=7.5$ & \cite{LandoltBornstein1998DielectricConstant}\tabularnewline
                            & $\varepsilon_{\mathrm{b}1}=\varepsilon_{\mathrm{s}2}=7.11$ & \cite{LandoltBornstein1998DielectricConstant}\tabularnewline
                            & $\varepsilon_{\mathrm{b}2}=6.46$ & \cite{LandoltBornstein1998DielectricConstant}\tabularnewline
       Energy of $\Gamma_{4}^{-}$-LO phonons & $\hbar\omega_{\mathrm{LO1}}=18.7\,\mathrm{meV}$ & \cite{KavoulakisBohrRadius}\tabularnewline
                            & $\hbar\omega_{\mathrm{LO2}}=87\,\mathrm{meV}$ & \cite{KavoulakisBohrRadius}\tabularnewline
     \hline
     \end{tabular}
\label{tab:MaterialParameters}
\end{table}

\subsection{Experimental methods}
\label{subsec:experiment}
{In this section, we present the experimental methods used in this paper for the observation of the
dark excitons in cuprous oxide.}
In short, the method {presented in
  Ref.~\cite{Farenbruch2020SHG}} was adjusted to optically activate 
the paraexcitons and 
to sensitively detect the resulting weak signals.

The paraexcitons are made allowed by applying a magnetic field, by which they gain an admixture of orthoexcitons through
the associated symmetry reduction. Because the symmetry for particular rotations around the magnetic field is still maintained,
the mixing occurs between states of the same symmetry class, which would be the same magnetic quantum number in the hydrogenlike
model. As a result of this coupling, the involved states, typically forming a two-level system, repel each other. Since the
coupling by the field is weak, the increase of the splitting between the states in the doublet can be well described by a
quadratic dependence on the magnetic field.

Despite of the hybridization of bright and dark states, the paraexciton oscillator
strength remains small, so that efforts had to be made to distinguish the corresponding lines from orthoexciton states: In
one-photon absorption the orthoexcitons with odd-symmetry envelopes, and among them mostly the $P$ excitons, clearly dominate
the spectra, also in magnetic field. Therefore, we turned to two-photon excitation, detected subsequently by the coherent
emission of photons at twice the frequency of the fundamental excitation laser, i.e., by second harmonic generation.

For the SHG experiments, we use 200 fs laser pulses with a spectral width of 10 meV. The 6 mm thick sample is cooled down to 1.4 K
being in superfluid Helium.
If allowed, the resulting spectra are typically dominated by orthoexcitons with even envelopes, but with much smaller oscillator
strength compared to the $P$ lines in one-photon absorption. This setting turned out to be sufficient to detect the paraexcitons
if they are in energy sufficiently separated from orthostates.

For the excited paraexciton states this separation may be too small,
so that we in addition chose experimental configurations with respect to the crystal orientation relative to the light propagation
as well as the polarization of the fundamental and second harmonic light, for which no SHG signal appears at zero
magnetic field, but appears only due to the field application. Doing so facilitated carving out the weak paraexciton signals up
to the principal quantum number $n = 6$.
{Further details of the experimental technique using optical second harmonic generation and its instrumental implementation
are given in Ref.~\cite{Farenbruch2020SHG}.}

\section{Results and Discussion}
\label{sec:ResultsDiscussion}

\subsection{Reversal of yellow $2S$ ortho- and paraexcitons}
\label{subsec:2S}
{In this section, we first briefly recapitulate the experimental
observation of the positions of the yellow 2S ortho- and paraexcitons,
presenting additional data not shown in Ref.~\cite{Farenbruch2020Dark}.
Note that we assign the labels \emph{green} and \emph{yellow}, as well
as the principal and angular quantum numbers in accordance with the
assignments given in Ref.~\cite{frankevenexcitonseries}.
We then present the underlying mechanisms.}
{Since the paraexciton is spin-flip forbidden in electrical dipole and quadrupole transition experiments,
we use a magnetic field to make them experimentally accessible. It is possible to include this magnetic field in the
Hamiltonian introduced in Sec.~\ref{subsec:Hamiltonian}, see for example
Refs.~\cite{frankmagnetoexcitonscuprousoxide,Rommel2018,Rommel2020SHG}.
In this work, we extrapolate
the experimental values to vanishing magnetic field and analyze those in the numerical calculations.
We therefore do not consider the magnetic field in the theory.}
\begin{figure}
\includegraphics[width=\columnwidth]{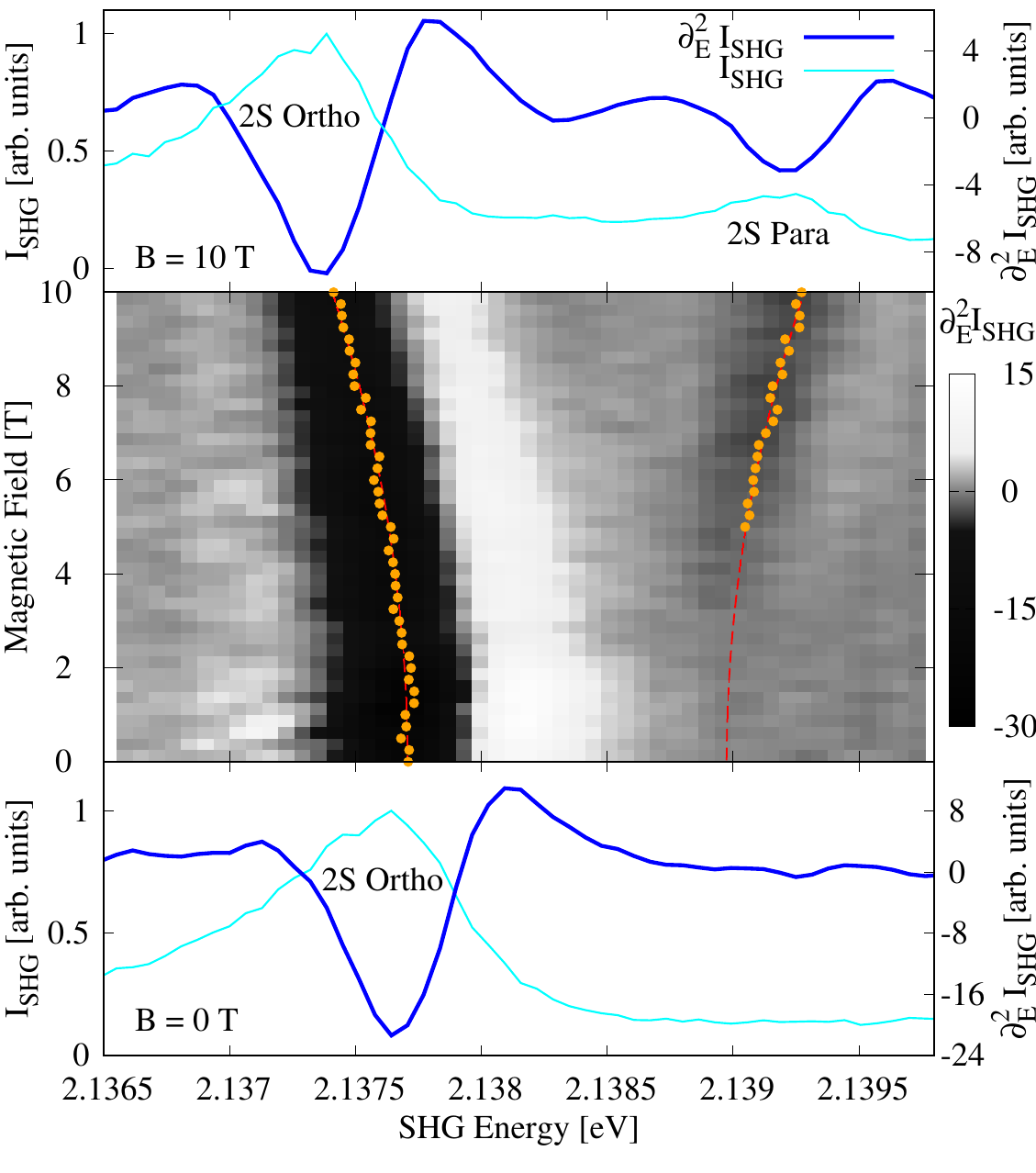}
\caption{
Experimental SHG spectrum of the yellow $2S$ ortho- and paraexcitons. The wave vector $\boldsymbol{K}$ is parallel to
the [111] axis and the magnetic field is applied in Voigt geometry along the [11$\overline{2}$] direction. The polarizations of the
incoming and outgoing light are parallel to the magnetic field. We show a contour plot of the second derivative
of the SHG intensity in gray scale. The positions of the ortho- and paraexcitons extracted by a Gaussian fit
to the SHG intensity are marked with orange dots. Using a quadratic
fit (red dashed line), we can extrapolate the energy of the
paraexciton to $E^{2S,\mathrm{para}}_{B = 0\,\mathrm {T}} = 2.13897\,\mathrm{eV}$ at vanishing magnetic field. An analogous fit to the orthoexciton
energies yields an energy $E^{2S,\mathrm{ortho}}_{B = 0\,\mathrm {T}} = 2.13771\,\mathrm{eV}$ at vanishing magnetic field.
The top and bottom panels show the SHG intensity and its second derivative at a magnetic field $B=10\,\mathrm{T}$ and $B=0\,\mathrm{T}$, respectively.}
\label{fig:Reversal2SParaOrtho}
\end{figure}

In the central panel of Fig~\ref{fig:Reversal2SParaOrtho},
we show a contour plot of the second derivatives of the SHG spectra of the $2S$ excitons.
The corresponding SHG spectra measured at a magnetic field of $10\,\mathrm{T}$ and $0\,\mathrm{T}$ are presented in
the top and bottom panel respectively.
Spectra are measured with a spectral resolution of $80\,\mathrm{\mu eV}$ from $0\,\mathrm{T}$ to $10\,\mathrm{T}$ in steps of $0.25\,\mathrm{T}$
in order to demonstrate
the spectral shift of the $2S$ para- and orthoexciton in a magnetic field. 
The wave vector is directed along the
SHG-allowed $[111]$ axis and the magnetic field is applied orthogonally to this in the [11$\overline{2}$] direction.
The polarization of the
incoming and outgoing light is parallel to the magnetic field, i.e. $\boldsymbol{E}^\mathrm{in} \parallel \boldsymbol{E}^\mathrm{out} \parallel$ [11$\overline{2}$].
This leads to a nonvanishing SHG signal of the paraexciton~\cite{Farenbruch2020Polarizations}.
It is much weaker than the intensity of the orthoexciton and only faintly becomes visible at about
$5\,\mathrm{T}$.
We therefore extrapolate the position of the paraexciton to zero magnetic field, using a quadratic fit. We obtain $E^{2S, \mathrm{para}}_{B = 0\,\mathrm {T}} = 2.13897\,\mathrm{eV}$ for
the paraexciton and $E^{2S, \mathrm{ortho}}_{B = 0\,\mathrm {T}} = 2.13771\,\mathrm{eV}$ for the orthoexciton.
We can therefore experimentally confirm one of the curious features of the yellow paraexciton series in cuprous oxide predicted by Schweiner~\emph{et al.}\
in Ref.~\cite{frankevenexcitonseries}, viz.\ the observation that the $2S$ paraexciton is located at a higher energy than the $2S$ orthoexciton.
{This shows that the experimentally observed behavior of the yellow excitons here is
qualitatively different from the hydrogenlike model in this respect.}
Farenbruch \emph{et al.}\ identify the origin of this
reversal in the influence of the green $1S$
exciton~\cite{Farenbruch2020Dark}.
{In the following, we want to corroborate this with a detailed
numerical analysis.}

\begin{figure}
\includegraphics[width=\columnwidth]{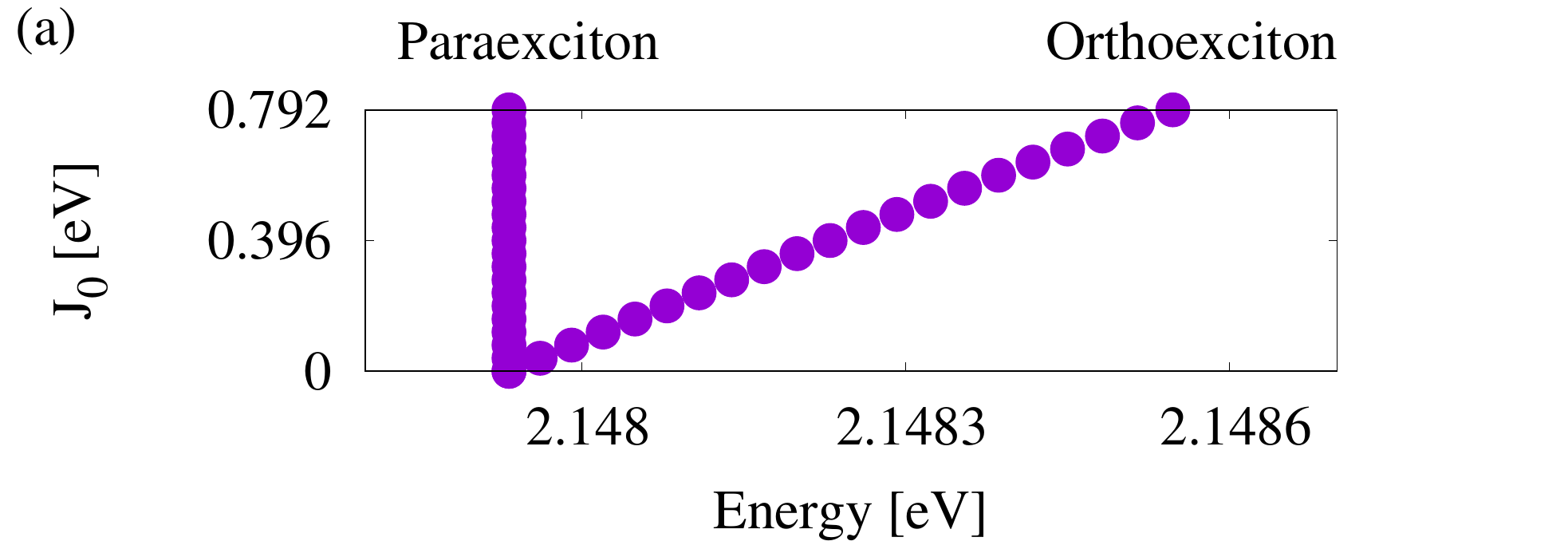}
\includegraphics[width=\columnwidth]{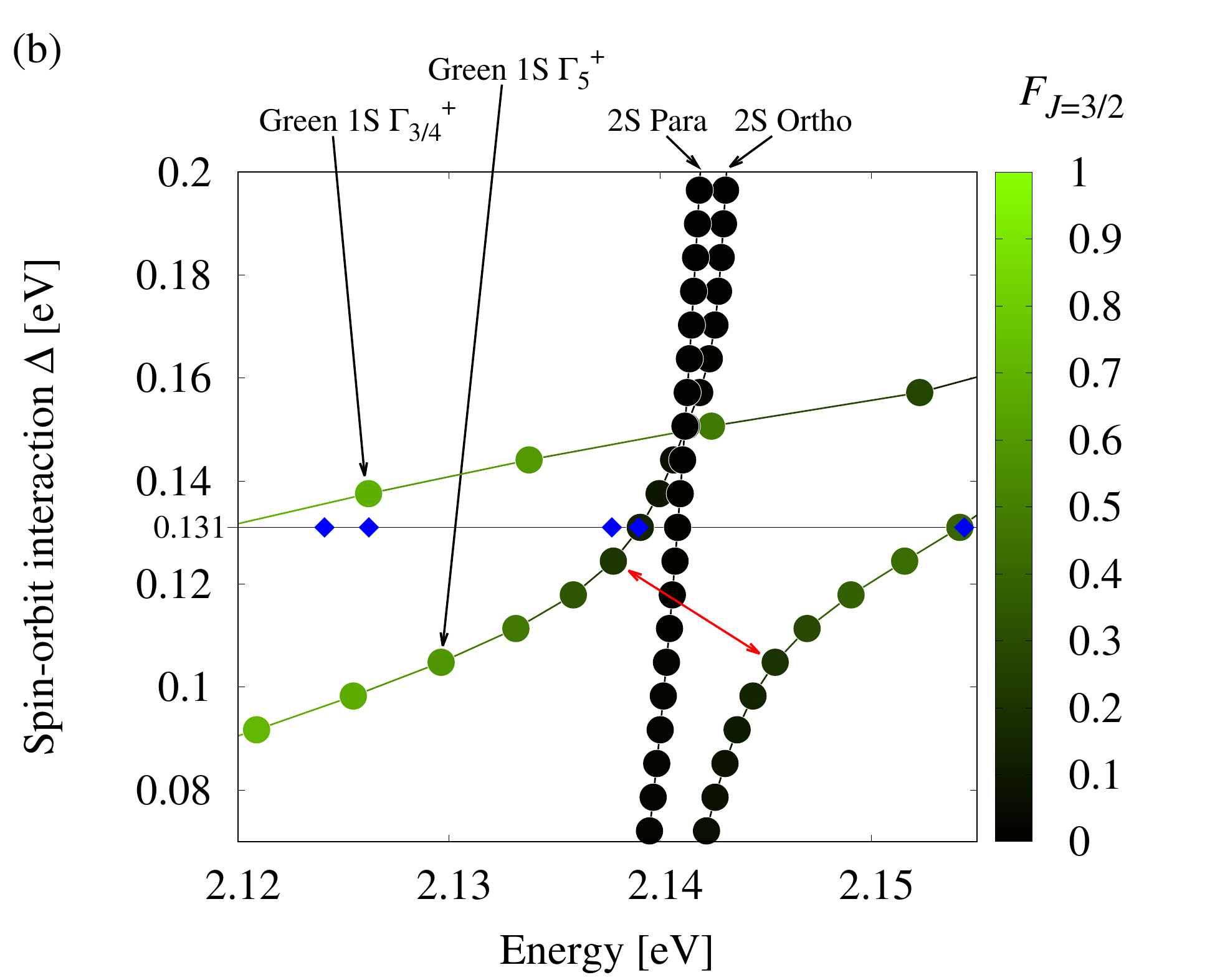}
\caption{(a) Splitting of the yellow $2S$ ortho- and paraexciton as a function of the strength of the exchange
interaction $J_0$ when only the yellow $J=1/2$ basis states are used for the diagonalization. In this case we observe that the
orthoexciton is lifted above the paraexciton as expected. The exchange interaction is fully switched on for $J_0 = 0.792\,\mathrm{eV}$~\cite{frankevenexcitonseries}.
This shows that the exchange of the positions of the para- and orthoexciton has to originate in the influence of
the green states.
In (b), we show the responsible avoided crossing in the spectrum near
the yellow $2S$ orthoexciton state as a function of the spin-orbit coupling $\Delta$. We added lines to help
guide the eyes. The
green admixture $F_{J=3/2}$ to the states is indicated by the color bar.
The horizontal line at $\Delta = 0.131\,\mathrm{eV}$ marks
the actual value of $\Delta$ in cuprous oxide. We can clearly see the avoided crossing between the yellow $2S$ orthoexciton
and the $\Gamma_5^+$ green $1S$ state, marked in red. It is this avoided crossing that places the $2S$ orthoexciton
below the $2S$ paraexciton. For larger values of $\Delta$, the influence of the green $1S$ state diminishes and the $2S$ orthoexciton
crosses the $2S$ paraexciton again and the usual order between those states is re-established. The experimental positions of
the exciton states are marked as blue diamonds. {We point out that the green $\Gamma_3^+$ and $\Gamma_4^+$ states
are degenerate in our model calculations, but show a small splitting in the experiment. This splitting is compatible with
the cubic symmetry of the crystal, but the effect is not captured by our Hamiltonian.}}
\label{fig:2SAvoidedCrossing}
\end{figure}

In Fig.~\ref{fig:2SAvoidedCrossing}(a) we show the exchange splitting for the yellow $2S$ state as a function of the parameter $J_0$,
with the green states removed from the spectrum. For this calculation, we only used states with $J=1/2$ in the basis. We see
that in this case, the exchange interaction lifts the orthoexciton above the paraexciton as predicted. This confirms that the
mixing with the $1S$ green orthoexciton is responsible for the surprising reversal, because without the green state, the reversal is absent.

For a better understanding, we calculate the positions of the yellow $2S$ and green $1S$ states as a function of the spin-orbit coupling,
revealing an avoided crossing. In Fig.~\ref{fig:2SAvoidedCrossing}(b) we show this avoided crossing between the yellow $2S$ and the green $1S$ orthoexciton.
The green admixture of the states given by the expectation value
\begin{align}
F_{J=3/2} = \langle \psi | P_{J=3/2}| \psi \rangle\,,
\end{align}
of the projection operator onto the $J=3/2$ Hilbert space for the exciton state $\psi$
is indicated by the color bar.
Using this green $J=3/2$ fraction we can identify the green states coming from the left-hand side and follow them through the crossing.
This avoided crossing was already noted in Ref.~\cite{frankevenexcitonseries}, but the implications for the relative placement of the $2S$
para- and orthoexcitons was not discussed. We can see that the avoided crossing leads to the yellow $2S$ orthoexciton being placed below the $2S$
paraexciton for the actual value of the spin-orbit coupling $\Delta = 0.131\,\mathrm{eV}$. For higher values at $\Delta \approx 0.15\,\mathrm{eV}$,
the ortho and paraexcitons cross each other again, when the influence of the green $\Gamma_5^+$ $1S$ state is small enough.
This further confirms and elucidates the influence of the mixing between the yellow and green series and its importance for a detailed understanding
of the yellow excitons.

\subsection{Dependence of the exchange splitting on the principal quantum number}
\label{subsec:PrincipalQuantumNumber}
\begin{figure}
\includegraphics[width=\columnwidth]{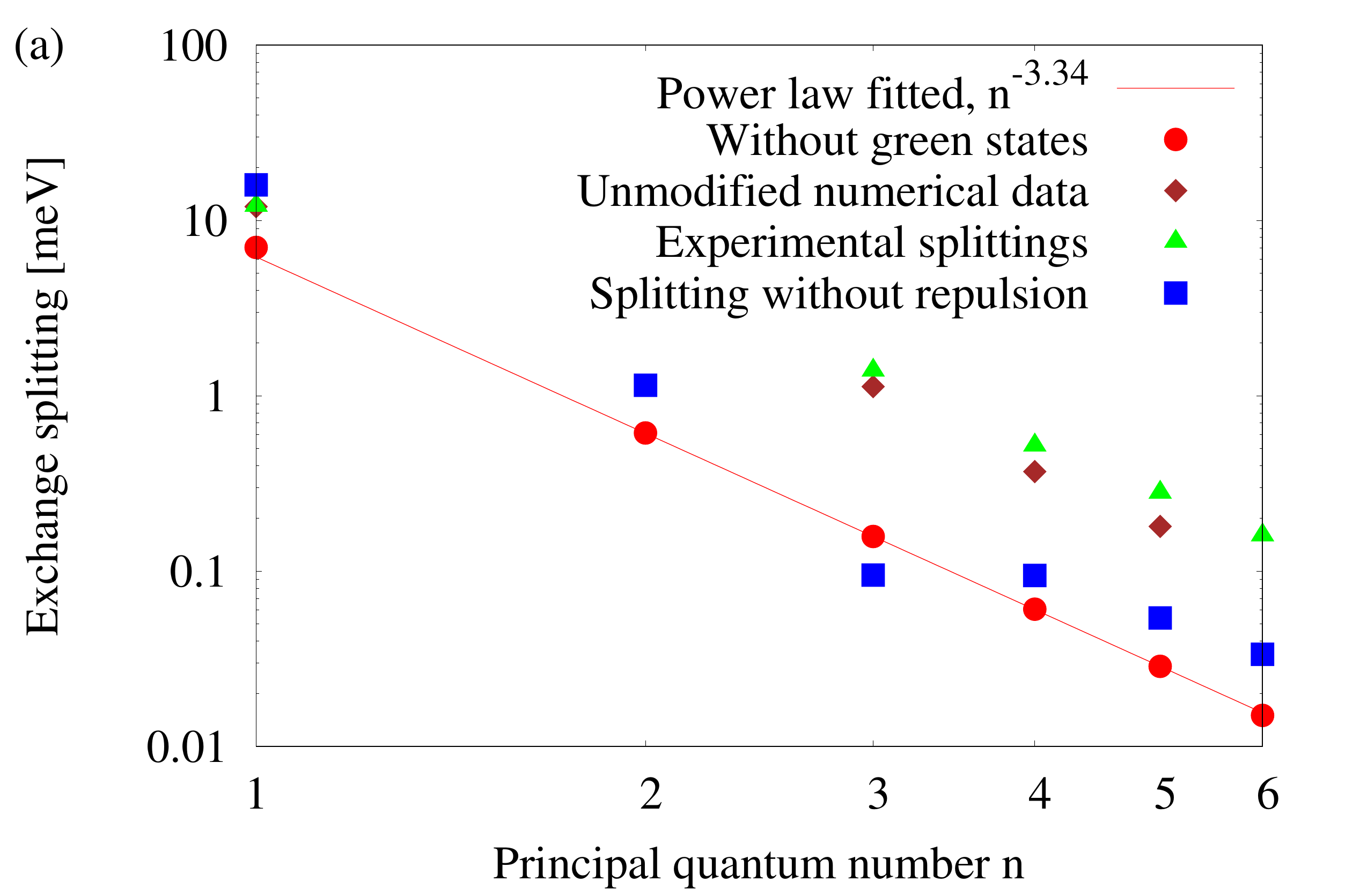}
\includegraphics[width=\columnwidth]{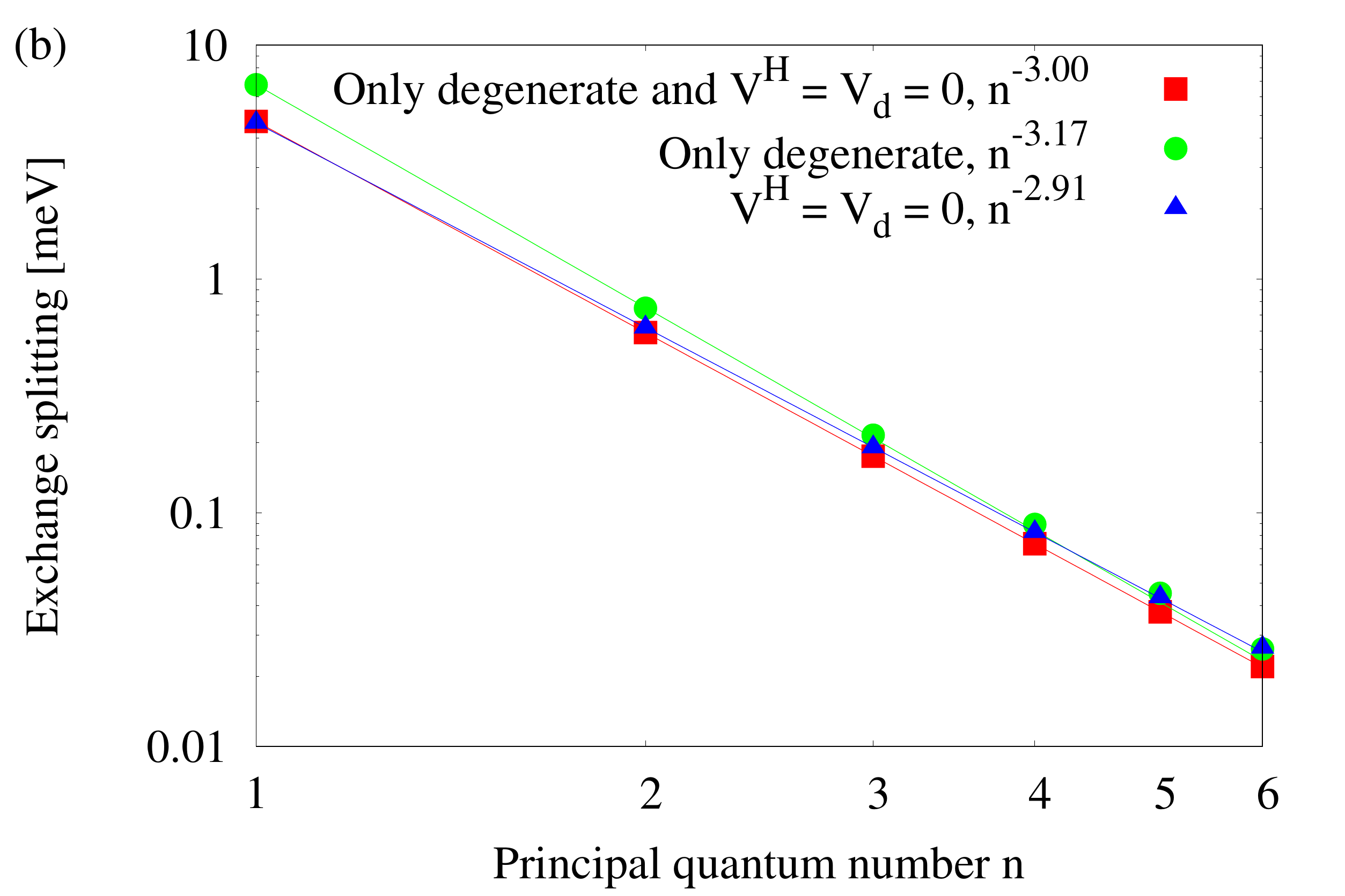}
\caption{Exchange splitting of the yellow excitons as a function of the principal quantum number $n$. To remove the influence
of the green $1S$ exciton, only basis states with $J=1/2$ were used. (a) Comparison of exchange splitting with (brown diamonds)
and without (red circles) the influence of the green $1S$ state. Unmodified numerical data were taken from Ref.~\cite{frankevenexcitonseries}.
We additionally show the experimental values (green triangles)
for reference. The blue squares show the splittings if the green state is included but the level repulsion
between the green and yellow states due to the $\delta$ terms in the central-cell corrections is removed.
(b) Exchange splitting as a function of $n$ for modified material parameters, again with only the yellow $J=1/2$ basis states.
We show data where we removed the influence of the Haken potential (blue triangles), data where we diagonalized the
exchange interaction only in the degenerate $S$ spaces, neglecting the coupling between different principal quantum numbers (green circles)
and data where we combined the previous two conditions (red squares). The fits show that only the combination of all modifications leads to the decrease
with the third power of the principal quantum number expected from the hydrogenlike model.}
\label{fig:ExchangeSplittingN}
\end{figure}

Since the removal of the mixing with the green $1S$ state restores the expected placement of ortho- and paraexcitons also in the
case of the yellow $2S$ state, it is a natural question whether the exchange splitting
decreases with the third power of the principal quantum number $n^{-3}$ as in the hydrogenlike model, Eq.~\eqref{eq:ExchangeFunctionN}.
In this section we want to investigate the exchange splitting of the yellow $S$ excitons as a function of $n$.
To remove the influence of the green $1S$ states, we only use the basis states with $J = 1/2$ belonging to the yellow series for the
calculations here. In Fig.~\ref{fig:ExchangeSplittingN}(a) we compare the numerical data for the full basis extracted from Table III in
Ref.~\cite{frankevenexcitonseries} with the exchange splitting if the influence of the green states is removed. We additionally show the actual
experimental values for reference. A fit of the form $\Delta E_\mathrm{exch}(n) = An^{B}$ reveals an exponent $B=-3.34$ still differing
from the expected $B=-3$ in the hydrogenlike model.

We identify two factors that explain this discrepancy. On the one hand, the Haken
potential modifies the dielectric constant for small radii. This leads to a change in the effective Bohr radius and thus
to a change in the value of the wavefunction at the origin. This disproportionally affects small quantum numbers, and thus
changes the dependency of the splitting on $n$.
On the other hand, the exchange interaction is not diagonal in the
principal quantum number, i.e., the $2S$ state also influences the $1S$ state and so on. Going back to Eq.~\eqref{eq:Exchange},
we see that the matrix elements do not necessarily vanish if the principal quantum numbers of the coupled states differ. This
also leads to a small but significant deviation from the $n^{-3}$ behavior.

We illustrate the effects of the different factors in Fig.~\ref{fig:ExchangeSplittingN}(b). {We find
that only if both of the factors discussed above are corrected for does the $n^{-3}$ behavior from the
hydrogenlike model emerge again.}

Interestingly, the removal of the green $1S$ state also has a significant effect on the absolute size of the splitting between
ortho and paraexcitons in the range of principal quantum numbers shown, as can be seen in Fig.~\ref{fig:ExchangeSplittingN}.

The most important effect accounting for
{this} is the level repulsion caused by the exchange interaction and $V_d$ matrix elements between the green orthoexciton and
the yellow $\Gamma_5^+$ states. The green $\Gamma_5^+$ exciton repels the yellow $\Gamma_5^+$ states, but the green $\Gamma_4^+$ and
$\Gamma_3^+$ states leave the yellow paraexcitons of symmetry $\Gamma_2^+$ unaffected. For yellow states energetically higher than the
green $1S$ state, this increases the splitting, whereas for those lower, it decreases it.
The blue squares in Fig.~\ref{fig:ExchangeSplittingN} (a) show the splittings of the yellow excitons
when this repulsion is removed. The resulting splittings in the yellow exciton series are far smaller than the experimental
values for $n \geq 4$ and more in line with the values when the green $1S$ state is removed completely, as can be seen by comparison
with the red circles. For the yellow $1S$ state, the effect is reversed and the removal of the level repulsion increases the
splitting instead.

\subsection{Splitting of yellow $1S$ state vs green $1S$ state}
\label{subsec:YellowVsGreen}
In this section we investigate the difference in the strength between the exchange splitting of the green $1S$ state
as compared to the yellow $1S$ state. We want to explain why the splitting of the green states is more than double that of
the yellow states.

Diagonalizing the yellow series alone without the green $J = 3/2$ basis states, we find that the splitting of the yellow $1S$ excitons is
approximately $7.02\,\mathrm{meV}$, which is even smaller than with the full basis. Diagonalizing the green states alone,
the splitting of the green $1S$ excitons is approximately $27.07\,\mathrm{meV}$. The discrepancy can therefore not be explained
by the mutual level shifts between the green $1S$ state and the yellow spectrum.

We again find that there are two factors which actually explain this difference. The first factor is the fact that the eigenvalues of the
operator $\boldsymbol{S}_\mathrm{e}\cdot\boldsymbol{S}_\mathrm{h}$ differ between the $J=1/2$ and $J=3/2$ Hilbert space.
According to the appendix of Ref.~\cite{frankevenexcitonseries}, the matrix elements in the basis~\eqref{eq:Basis} are given by
\begin{align}
M = &\left\langle \Pi'\left|\boldsymbol{S}_{\mathrm{e}}\cdot\boldsymbol{S}_{\mathrm{h}}\,\delta\left(\boldsymbol{r}\right)\right|\Pi\right\rangle\nonumber\\
 =  &\: \delta_{L'0}\delta_{L0}\delta_{F_{t}F'_{t}}\delta_{M_{F_{t}}M'_{F_{t}}}\:\frac{3}{2\pi}\left(-1\right)^{F_\mathrm{t}+F'+F+J+J'}\nonumber \\
\times &\: \left[\left(2F+1\right)\left(2F'+1\right)\left(2J+1\right)\left(2J'+1\right)\right]^{\frac{1}{2}}\nonumber \\[1ex]
\times &\: \left\{ \begin{array}{ccc}
F' & F & 1\\
\frac{1}{2} & \frac{1}{2} & F_\mathrm{t}
\end{array}\right\} \left\{ \begin{array}{ccc}
F & F' & 1\\
J' & J & 0
\end{array}\right\} \left\{ \begin{array}{ccc}
\frac{1}{2} & J' & 1\\
J & \frac{1}{2} & 1
\end{array}\right\}\,.
\label{eq:MatrixElementExchange}
\end{align}
Fixing either $J=J'=1/2$ or $J=J'=3/2$, the operator is already diagonal in the given basis.
We can evaluate the matrix elements for the yellow and green series and $L = L' = 0$. For $J=J'=1/2$, we calculate $M = 1/4\pi$ with $F_\mathrm{t} = 0$
and $M=-1/12\pi$ with $F_\mathrm{t} = 1$. For $J=J'=3/2$ it is $M=1/4\pi$ when $F_\mathrm{t} = 2$ and $M = -5/12\pi$ when $F_\mathrm{t} = 1$.
Note that the exchange interaction~\eqref{eq:Exchange} contains
this operator with reversed sign. The exchange interaction therefore lifts the $F_\mathrm{t} = 1$ states above the others in both
the yellow and green series as depicted in Fig.~\ref{fig:ExchangeVSSO}.
We thus find that the splitting in the eigenvalues for $J=3/2$ is $\Delta M_\mathrm{green} = 2/3\pi$ and consequently exactly double the
splitting for $J=1/2$, which is $\Delta M_\mathrm{yellow} = 1/3\pi$.

These calculations account for part of the difference between the yellow and green splitting. A factor of approximately 1.93 between the green and yellow splitting
remains to be explained. Because of the $\delta$ term, the exchange splitting is proportional to $|\psi(0)|^2$, which in turn is proportional to the inverse third
power of the Bohr radius $a_\mathrm{B}^{-3}$. Since the Bohr radius is proportional to the reduced mass $\mu$, it follows that
$|\psi(0)|^2 \sim \mu^3$. This seems to be the explanation for the factor $1.93$, as explained in the following. The reduced mass
is proportional to the Rydberg energy in a hydrogenlike system. To approximate the latter, we calculated the binding energy of the yellow and green $1S$ states
while varying the exact form of the potential. The results are listed in Table~\ref{tab:RydbergEnergies}.
\begin{table}[t!]
\caption{Energies of the lowest yellow and green $1S$ excitons for different choices of the parameters in the
central-cell corrections with the exchange interaction removed.
For the yellow values, we only diagonalized the $J=1/2$ Hilbert space, and for the green values only the $J=3/2$ Hilbert space.
For the gap energies we used $E_\mathrm{gap,yellow} = 2.17208\,\mathrm{eV}$ and $E_\mathrm{gap,green} = 2.30308\,\mathrm{eV}$.}
\begin{tabular}{ccccc}
 series & $V_d $ & $V^{\mathrm{H}}$ & $E_{1S}\,$[eV] & $E_\mathrm{Ryd}\,$[meV] \\
\hline
yellow&  on& on & 2.059  &  112.8 \\
yellow&  on& off &  2.076  & \phantom{1}95.9   \\ 
yellow&  off& off & 2.086  & \phantom{1}86.1   \\ 
green&  on& on &  2.153 &  150.5  \\
green& on& off &  2.179 &  124.1 \\
green&  off& off & 2.198 &   105.3
\end{tabular}
\label{tab:RydbergEnergies}
\end{table}
Based on these data, we can estimate the ratio of the reduced masses of the green and yellow
$1S$ states with the ratio of the binding energies. Since the latter are not only affected by the Coulomb interaction, but also by
the additional terms $V^{\mathrm{H}}$ and $V_d$ in Eq.~\eqref{eq:CCCHaken}, we need to correct for those.
Using the values where the central-cell corrections are removed completely, we get
\begin{align}
\left(\frac{\mu_\mathrm{green}}{\mu_\mathrm{yellow}}\right)^3 & \approx \bigg(\frac{E^{\mathrm{green}}_\mathrm{Ryd}}{E^{\mathrm{yellow}}_\mathrm{Ryd}}\bigg)^3 \nonumber\\
 &\approx \left(\frac{105.3\,\mathrm{meV}}{86.1\,\mathrm{meV}}\right)^3 \approx 1.22^3 \approx 1.82\,.
\end{align}
This is in good agreement with the factor of 1.93.
The explanation for the different strengths of the exchange splitting in the yellow and green $1S$ exciton states
therefore is on the one hand the factor two due to the operator $\boldsymbol{S}_\mathrm{e}\cdot\boldsymbol{S}_\mathrm{h}$
for $J=1/2$ and $J=3/2$ and on the other hand the difference in the reduced mass $\mu$ for the yellow and green $1S$ exciton.

\section{Summary and conclusion}
\label{sec:Conclusion}
Experimental investigations into the paraexciton series of yellow excitons in Cu$_2$O and corresponding exchange splittings reveal
a number of ways in which a simple hydrogenlike model is insufficient. In this manuscript, we numerically
investigated spectra with modified material parameters and thus gained experimentally inaccessibly insights.
We used this to interpret the experimental findings in Ref.~\cite{Farenbruch2020Dark} and identify their roots
in the properties of the system.

We first investigated the reversal of the yellow $2S$ para- and orthoexcitons. Farenbruch \emph{et al.}~\cite{Farenbruch2020Dark}
identified the mixture with the $1S$ green orthoexciton as the origin of the reversal. We were able to corroborate this explanation
with detailed calculations. We show that the orthoexciton is lifted above the paraexciton if the influence of the green excitons
is removed in the simulation. Varying the spin-orbit coupling reveals an avoided crossing between the yellow $2S$ orthoexciton and the green
$1S$ exciton which explains the placement of the orthoexciton below the paraexciton. {We were thus able to show how the
coupling of the yellow and green series leads to a behavior that qualitatively differs from the hydrogenlike approximation,
underscoring its importance for the understanding of the yellow exciton series.}

Removing the influence of the green states, the expected order of states is restored. In this case, does the exchange splitting
decrease with the third power of the principal quantum number $n$? Our calculations show that this is not exactly the case.
We identify two reasons for this. First of all, the Haken potential changes the dielectric constant for small radii, which
influences the wavefunction at the origin and therefore the splitting. A simplified treatment of the exchange splitting also
overlooks the second factor, which is the coupling between $S$ states of different principal quantum number by the exchange
splitting itself. A systematic analysis shows that these two factors account for the discrepancy from the $n^{-3}$ behavior.

We {conclude by studying} the origin of the large difference between the exchange splittings of the
yellow and green $1S$
states. Farenbruch \emph{et al.}~\cite{Farenbruch2020Dark} confirmed the prediction by
Schweiner \emph{et al.}~\cite{frankevenexcitonseries}
that the splitting of the green $1S$ exciton is over $30\,\mathrm{meV}$ and therefore
{about two and a half times} the splitting of
about $12\,\mathrm{meV}$ for the yellow $1S$ state. We also identified two reasons to account for this. The first is the difference
in the matrix elements of $\boldsymbol{S}_\mathrm{e}\cdot\boldsymbol{S}_\mathrm{h}\delta(\boldsymbol{r})$ for $J=1/2$ and $J=3/2$.
Since the exchange
splitting depends upon the relative orientation of the electron and hole spins, different values of the effective hole spin
$\boldsymbol{J} = \boldsymbol{I} + \boldsymbol{S}_\mathrm{h}$ lead to different strengths of the exchange splitting.
The second is the
difference in the reduced masses between the yellow and green $1S$ states. The reduced mass of the green $1S$ state is significantly
higher than the
reduced mass of the yellow $1S$ state as revealed by a detailed analysis of the Rydberg energy when correcting for the influence of
short distance terms in
the Hamiltonian. This leads to a higher value of the wavefunction at the origin and a corresponding increase of the {exchange}
splitting.

\acknowledgments
The theoretical studies at University of Stuttgart were supported by
Deutsche Forschungsgemeinschaft (Grant No.\ MA1639/13-1).
The experimental studies at the TU Dortmund University were supported by the Deutsche
Forschungsgemeinschaft through the International Collaborative
Research Centre TRR 160 (Projects No. A8 and C8).
We also acknowledge the support by the project AS 459/1-3.
We thank Frank Schweiner for his contributions.


%

\end{document}